\documentclass[sigconf, nonacm]{acmart}

\newcommand\vldbyear{2026}
\newcommand\vldbworkshop{ADS 2026: The Joint Workshop on Agentic Data Systems and Data-Centric AI (The 1st ADS \& 3rd DATAI)}
\newcommand\vldbauthors{\authors}
\newcommand\vldbtitle{\shorttitle}

\newcommand\vldbpagestyle{plain}

\usepackage{xcolor}
\usepackage{enumitem}
\usepackage{xspace}

\newcommand{\system}{chart-plot\xspace}

\begin{document}

\title{Demonstrating chart-plot: Closing the Last Mile of Academic Chart Generation}

\author{Yinghao Tang}
\affiliation{%
  \institution{State Key Lab of CAD\&CG, Zhejiang University}
  \city{Hangzhou}
  \country{China}}
\email{yinghaotang@zju.edu.cn}

\author{Yupeng Xie}
\affiliation{%
  \institution{HKUST(GZ)}
  \city{Guangzhou}
  \country{China}}
\email{yxie740@connect.hkust-gz.edu.cn}

\author{Yingchaojie Feng}
\affiliation{%
  \institution{National University of Singapore}
  \city{Singapore}
  \country{Singapore}}
\email{feng.y@nus.edu.sg}

\author{Jiale Lao}
\affiliation{%
  \institution{Cornell University}
  \city{Ithaca}
  \state{NY}
  \country{USA}}
\email{jiale@cs.cornell.edu}

\author{Tingfeng Lan}
\affiliation{%
  \institution{University of Virginia}
  \city{Charlottesville}
  \state{VA}
  \country{USA}}
\email{tingfeng@virginia.edu}

\author{Wei Chen}
\affiliation{%
  \institution{State Key Lab of CAD\&CG, Zhejiang University}
  \city{Hangzhou}
  \country{China}}
\email{chenvis@zju.edu.cn}

\renewcommand{\shortauthors}{Tang et al.}

\begin{abstract}
Large language models can translate a researcher's intent into runnable
\texttt{matplotlib} code, yet the resulting chart rarely lands in a paper
without multiple rounds of manual revision. We argue that the open problem
is not chart \emph{code} generation but chart \emph{publication}: making
the output look like a top-venue figure, survive the target layout, and
respond to precise author edits. We present \textbf{\system{}}, an
agentic harness that closes this last mile through three components:
(1)~a style-aware code generator conditioned on a textual style skill
distilled from accepted figures at the target venue, (2)~a
deployment-aware render loop that compiles the chart inside the target
\LaTeX{} context and revises until layout constraints are met, and
(3)~a structured edit layer that exposes every chart element as a
directly manipulable handle. We report early results on three chart-type
case studies (grouped bar, scaling line, paired distributions) and a
small user study.
\end{abstract}

\maketitle

\pagestyle{\vldbpagestyle}
\begingroup\small\noindent\raggedright\textbf{VLDB Workshop Reference Format:}\\
\vldbauthors. \vldbtitle. VLDB \vldbyear\ Workshop: \vldbworkshop.\\
\endgroup
\begingroup
\renewcommand\thefootnote{}\footnote{\noindent
This work is licensed under the Creative Commons BY-NC-ND 4.0 International License. Visit \url{https://creativecommons.org/licenses/by-nc-nd/4.0/} to view a copy of this license. For any use beyond those covered by this license, obtain permission by emailing \href{mailto:info@vldb.org}{info@vldb.org}. Copyright is held by the owner/author(s). Publication rights licensed to the VLDB Endowment. \\
\raggedright Proceedings of the VLDB Endowment.
ISSN 2150-8097. \\
}\addtocounter{footnote}{-1}\endgroup

\vspace{.3cm}
\begingroup\small\noindent\raggedright\textbf{VLDB Workshop Artifact Availability:}\\
The source code, data, and other artifacts will be released as open source upon paper acceptance.
\endgroup

\section{Introduction}
\label{sec:intro}

A modern paper has dozens of figures, many of them charts: ablation
bar plots, scaling curves, scatter plots of latent embeddings.
Large language models can already translate a researcher's intent
and a CSV into runnable \texttt{matplotlib} code in
seconds. Yet the chart that comes out of the
model still rarely makes it into the paper without two or three
rounds of manual edits.

\begin{figure}[t]
  \centering
  \includegraphics[width=\linewidth]{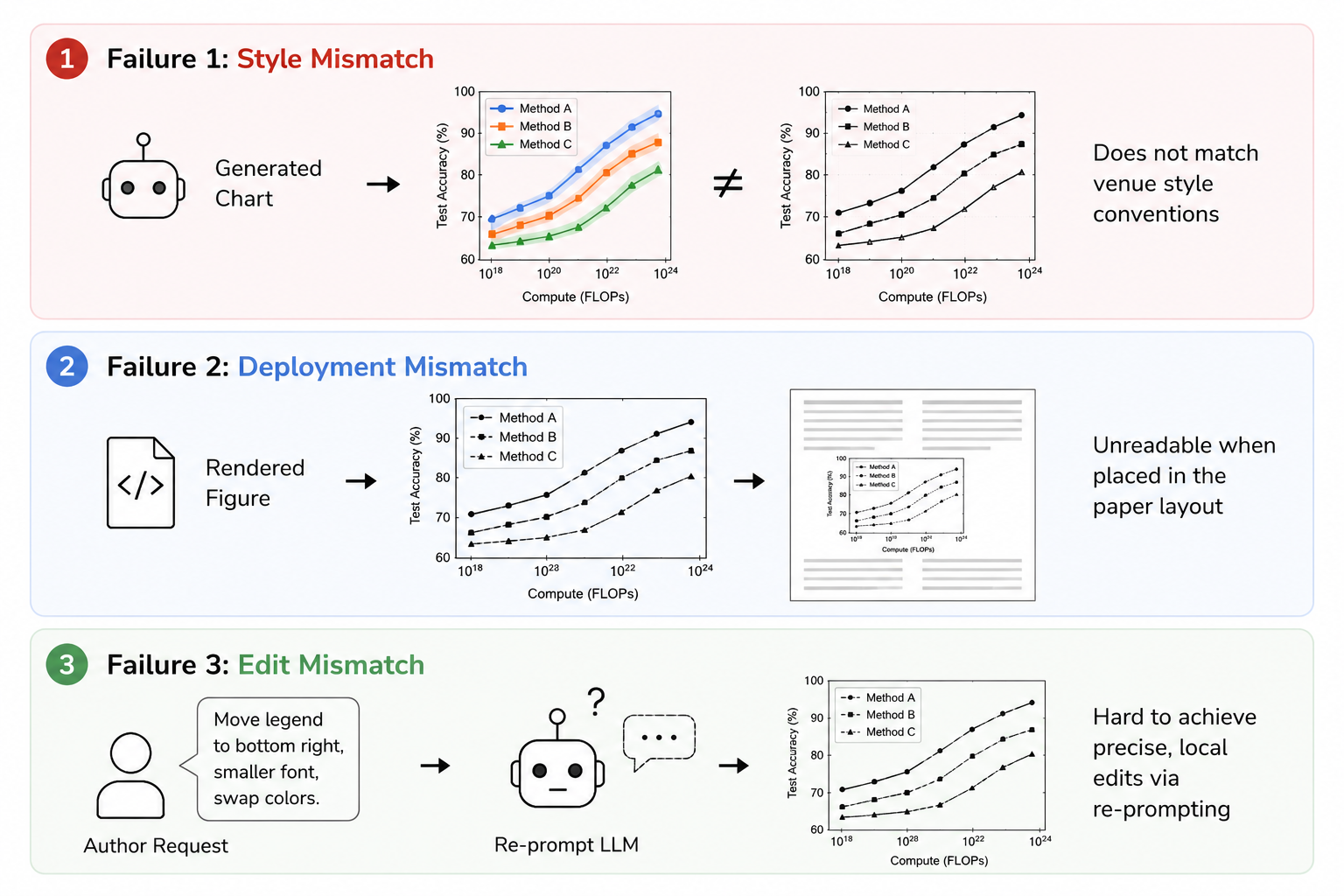}
  \caption{The last mile. A generated chart becomes publication-ready
  only after three paper-facing concerns are resolved together:
  matching the visual conventions of the target venue, surviving
  column-width deployment, and supporting precise author edits.}
  \label{fig:teaser}
\end{figure}

We argue that the open problem is not chart \emph{code} generation;
as model capabilities advance, existing systems handle that step
increasingly well~\cite{lida2023, chartllama2023, matplotagent2024,
vispath2025, chartcoder2025, tang2026vividoc, tang2026igenbench}. The open problem is chart
\emph{publication}: turning a generated chart into a figure that
looks at home in a top-venue paper, survives the target layout, and
can be fixed with a small targeted edit when the model gets it
almost right. We call this gap the \textbf{last mile of academic
chart generation} and identify three concrete failure modes
(\autoref{fig:teaser}).

\paragraph{Failure 1: style mismatch.} Top-venue charts follow
community conventions. AI papers favour low-saturation palettes,
sans-serif labels, and confidence ribbons; systems papers favour
serif labels, hatched bars, and explicit error bars. A naively
generated chart is plausible but visually off-style.

\paragraph{Failure 2: deployment mismatch.} A chart rendered at
\texttt{matplotlib}'s default 6.4$\times$4.8 inches looks fine in
isolation. Dropped into a two-column ACM template at 3.3-inch
column width, tick labels become unreadable, legends overlap data,
and titles clip. The figure is correct only once it compiles inside
the paper.

\paragraph{Failure 3: edit mismatch.} Once a chart is mostly right,
authors want small geometric edits: move a legend, shrink an
annotation, swap two colours. Re-prompting an LLM is unreliable
here because the model cannot localise a few-pixel change from a
text instruction alone.

\paragraph{Our approach: a harness.} We close these gaps with a
\textbf{harness}~\cite{gu2026harness, ning2026code}: a
deterministic scaffold around a generative LLM that injects
publication context the model cannot infer from a prompt alone,
such as the column width of the target template or the venue's
visual conventions. \system{} organises this scaffold across a
spectrum from full automation to author control:
(1) a \emph{style} layer conditions generation on the venue's
visual conventions;
(2) a \emph{deployment} layer revises the figure automatically
until it fits the target \LaTeX{} template; and
(3) an \emph{edit} layer hands the author direct handles on
every figure element when geometric judgment is required.
The result is a closed loop that ends only when the figure is
publication-ready.

\paragraph{Contributions.} We present \textbf{\system{}}, an
agentic harness that closes the last mile through three components,
each targeting one failure mode.

\begin{itemize}[leftmargin=*, itemsep=2pt]
  \item A \textbf{style-aware code generator} that conditions on a
  curated corpus of accepted figures, distilled into a textual
  style skill that is injected at generation time
  (\S\ref{sec:method:style}).
  \item A \textbf{deployment-aware render loop} that compiles the
  chart inside the target \LaTeX{} context, detects layout failures,
  and revises the code until constraints are met
  (\S\ref{sec:method:deploy}).
  \item A \textbf{structured edit layer} that exposes every
  \texttt{matplotlib} element as a directly manipulable handle,
  mapping author gestures back to code mutations
  (\S\ref{sec:method:edit}).
\end{itemize}

Together, the three components form a harness that slots into any
LLM-backed analysis workflow as the final reporting step: the
researcher provides data and intent, and \system{} delivers a
publication-ready figure.

\section{Related Work}
\label{sec:related}

\subsection{Agent Harness Engineering}
\label{sec:related:harness}

Early tool-using agents introduced the basic loop of reasoning, action, and observation, making external tools and feedback part of model behavior rather than post-hoc additions~\cite{yao2023react,schick2023toolformer}. Building on these primitives, recent agent systems increasingly move reliability into the harness: the infrastructure that manages execution, tool interfaces, context, orchestration, verification, and human control. Multi-agent and workflow frameworks make orchestration explicit~\cite{wu2023autogen,hong2024metagpt,li2026deepeye}, while coding and web agents couple models with controlled execution environments, tool protocols, and benchmark feedback~\cite{yang2024sweagent,wang2025openhands,zhou2024webarena,jimenez2024swebench,tang2025scorpio}. Recent harness-specific work makes this shift explicit by arguing that long-horizon agent performance is often bounded by the wrapper around the model, and by studying harness taxonomies, harness scaling, executable code harnesses, and automated harness optimization~\cite{li2026agentharness,gu2026harness,ning2026code}. \system{} follows this harness view, but specializes it to academic chart publication: its context injection, execution loop, verification checks, and author controls are organized around the \emph{last mile} from generated chart code to a paper-ready figure.

\subsection{Publication-Aware Chart Generation}
\label{sec:related:publication}

LLM-based visualization systems have made chart generation increasingly executable, multimodal, and self-correcting. Early systems translate data and intent into visualization code through multi-stage generation pipelines~\cite{lida2023}, while chart-specific models and benchmarks improve chart generation, chart-to-code recovery, and aesthetic evaluation through instruction tuning, multimodal corpora, and executable program synthesis~\cite{chartllama2023,zadeh2024text2chart31,chartcoder2025,xie2025visjudge,tang2026igenbench}. More recent agentic systems add execution, debugging, visual critique, specialized critics, and multi-path feedback loops to improve generated visualization scripts~\cite{matplotagent2024,vispath2025,visshepherd2025,wang2026ggplotagent,xie2024haichart,datalab2024}. In parallel, visualization authoring work studies how users refine communicative figures through design constraints, direct manipulation, code links, mixed-initiative interfaces, and human-agent collaboration~\cite{draco2018,zong2020lyra,wang2025data,vaithilingam2024dynavis,chen2025chartmark,tang2026vividoc,tang2026vividocdemo}; scientific figure tools further emphasize reproducible post-generation editing and image- or vector-level refinement~\cite{gerum2019pylustrator,chartreformer2024,zhu2026autofigure}. These lines improve chart code, standalone visual quality, or interactive editing, but they usually do not treat the target paper as the execution environment. \system{} instead makes publication context first-class: venue style, \LaTeX{} layout constraints, structured chart layout, and author edits are joined in one harness so that generation ends with a reproducible, publication-ready figure.
\section{System Overview}
\label{sec:system}

\begin{figure*}[t]
  \centering
  \includegraphics[width=0.9\textwidth]{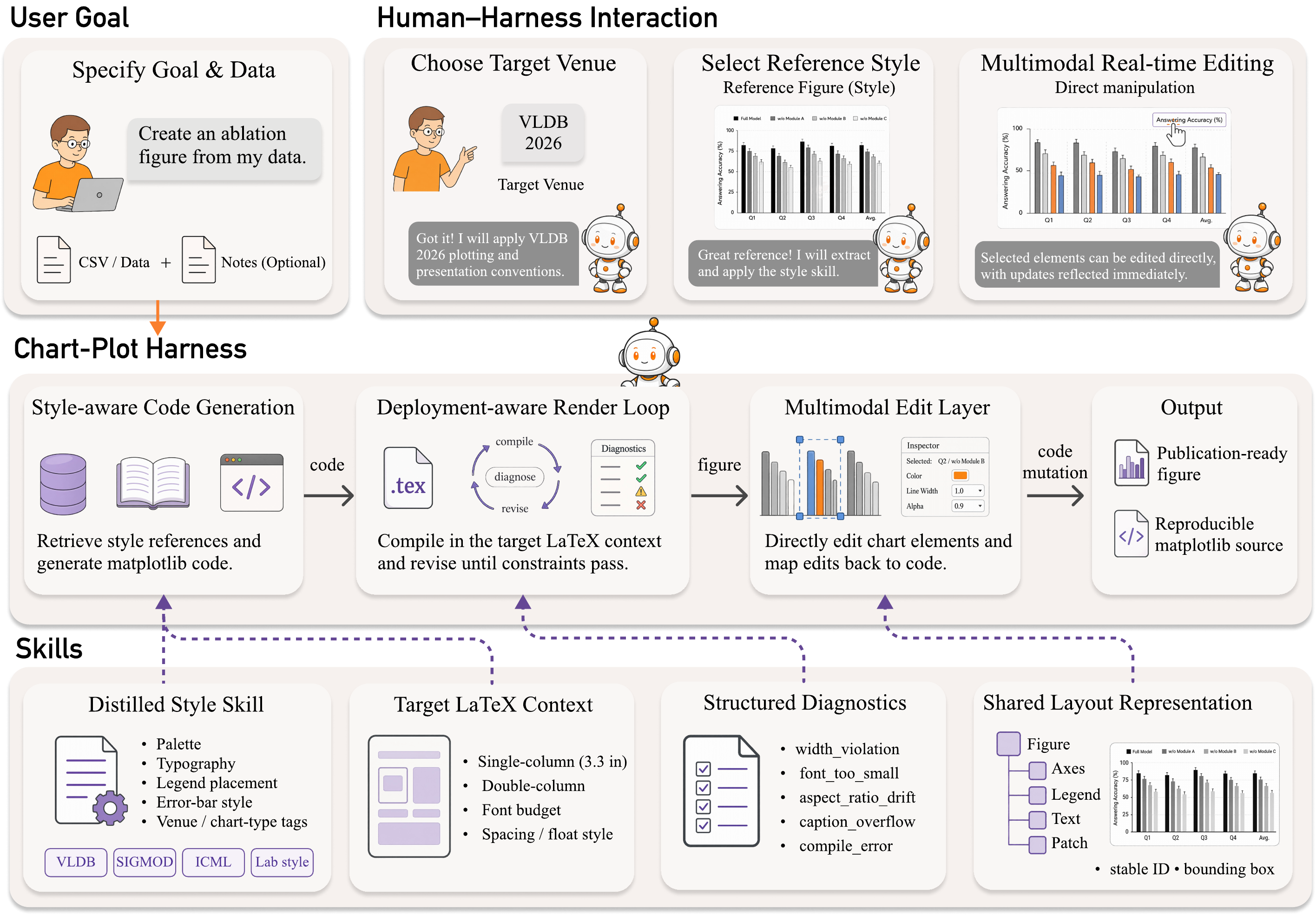}
  \Description{A three-tier architecture diagram of the chart-plot
  harness. The top tier shows the author's interaction with the
  system: specifying a goal and data, choosing a target venue,
  selecting a reference style, and editing the figure in real time.
  The middle tier shows the harness stages: style-aware code
  generation, a deployment-aware render loop, a multimodal edit
  layer, and the final output of a publication-ready figure with
  reproducible source. The bottom tier shows the shared skills and
  representations that the stages draw on: the distilled style
  skill, the target LaTeX context, structured diagnostics, and a
  shared layout representation.}
  \caption{The \system{} architecture. The author specifies a
  goal and data, selects a target venue and an optional reference
  style, and may edit the result through direct manipulation. The
  harness then runs three stages: style-aware code generation
  loads the distilled style skill and writes \texttt{matplotlib}
  code (\S\ref{sec:method:style}); the deployment-aware render
  loop compiles the figure inside the target \LaTeX{} context and
  revises the code until layout constraints are satisfied
  (\S\ref{sec:method:deploy}); and the multimodal edit layer maps
  direct manipulations back to code mutations
  (\S\ref{sec:method:edit}). A shared layout representation with
  stable element identifiers and bounding boxes is reused across
  the deployment and edit stages.}
  \label{fig:pipeline}
\end{figure*}

\autoref{fig:pipeline} shows the \system{} architecture. The
system takes four inputs: a data table, an intent string, a target
venue, and a target \LaTeX{} template; the author may optionally
select a reference style to steer the venue's visual conventions.
It returns a publication-ready figure together with the reproducible
\texttt{matplotlib} source that produced it.

\paragraph{Design philosophy.} We frame \system{} as a
\emph{harness}: infrastructure that channels the raw generative
power of a code LLM into a publication-ready artifact by learning
the gold standard of each target scenario. The harness operates
at three levels, arranged along an automation spectrum.

\begin{enumerate}[leftmargin=*, itemsep=2pt]
  \item \textbf{Style} (recommended, user-selectable). The harness
  learns visual conventions from a curated corpus of accepted
  figures, compiled into a textual style skill that is injected
  into the codegen system prompt at generation time. The author
  can swap in a different skill per venue or per lab
  (\S\ref{sec:method:style}).
  \item \textbf{Deployment} (fully automated). The harness learns
  the layout rules of the target template (column width, font
  budget, spacing) and enforces them through a compile-and-revise
  loop. No author intervention is needed unless the constraints are
  fundamentally unsatisfiable (\S\ref{sec:method:deploy}).
  \item \textbf{Edit} (author-controlled). For preferences that no
  model can anticipate, the harness exposes click-targets over the
  rendered figure: selecting an element opens a contextual
  inspector with controls specific to its kind (legend location,
  font size, color, line style, etc.). Each control maps to a
  concrete code mutation (\S\ref{sec:method:edit}).
\end{enumerate}

\paragraph{Shared representation.} After code generation, the
system executes the \texttt{matplotlib} program in a sandbox and
walks the resulting \texttt{Figure} object tree. Every element
(axis, legend, text, patch, line) receives a stable identifier and
a bounding box. This layout representation is shared across the
deployment loop and the edit layer, so automated fixes and manual
edits compose without conflict.

\system{} closes the generation loop: it checks the output
against learned standards, revises when constraints fail, and
hands control to the author when the remaining gap is a matter of
personal preference.

\section{Method}
\label{sec:method}

\subsection{Style-aware code generation}
\label{sec:method:style}

Top-venue charts follow community-specific conventions (e.g.\
low-saturation palettes at AI venues, hatched bars at systems
venues), and a single generation prompt cannot capture these norms
in sufficient detail. We treat style as a corpus distillation
problem.

Our corpus is built by extracting figures from accepted papers at
target venues. Each PDF is parsed into per-figure PNGs paired with
their captions; the candidate figures are then human-curated for
visual quality and chart-likeness, and the surviving entries are
tagged with venue and chart type (ablation, scaling curve,
training curve, etc.). The corpus grows monotonically as new
venues accept papers.

The corpus is distilled into a textual \emph{skill}: a markdown
summary of the recurring visual conventions (palette, font sizing,
legend placement, error-bar style) that is loaded into the codegen
LLM's context at generation time. We follow the published skill
convention~\cite{anthropic_skills} of attaching a self-contained,
version-controlled instruction file alongside an agent's tool
surface. The artifact is a single file the author can read,
version, or override per venue.

\begin{figure*}[t]
  \centering
  \begin{minipage}{0.495\textwidth}\centering
    \includegraphics[width=\linewidth]{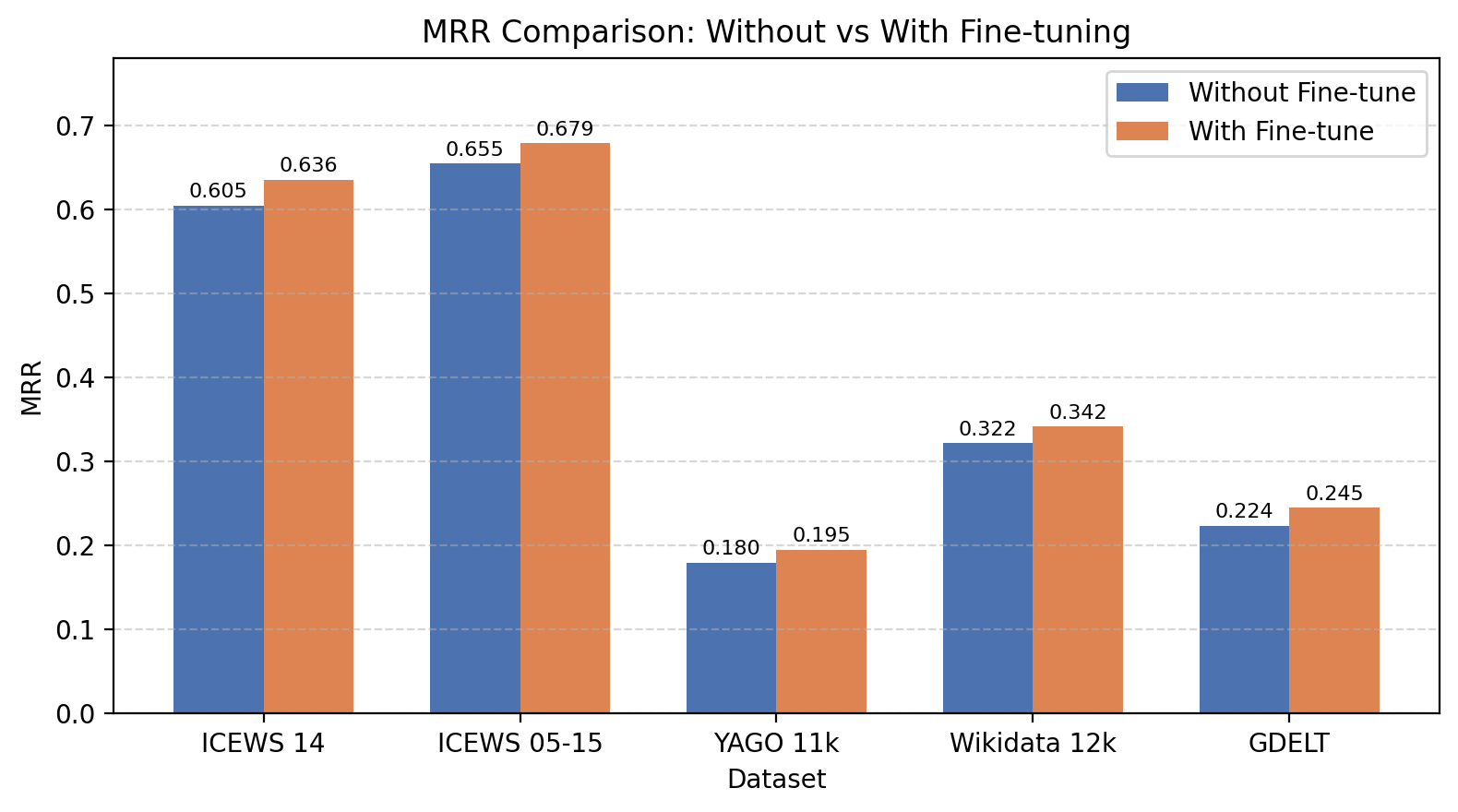}
  \end{minipage}\hfill
  \begin{minipage}{0.495\textwidth}\centering
    \includegraphics[width=\linewidth]{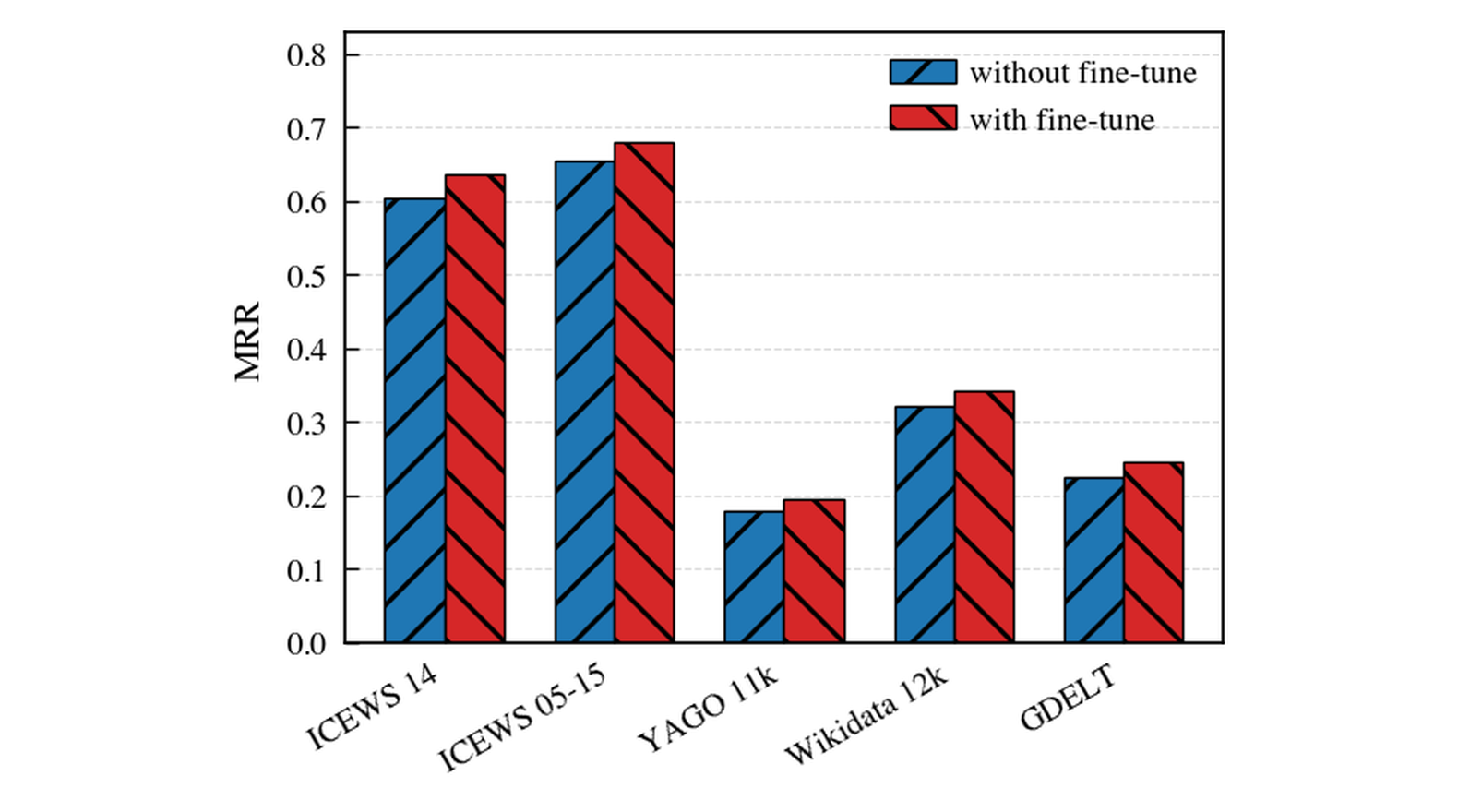}
  \end{minipage}
  \\[4pt]
  \begin{minipage}{0.495\textwidth}\centering
    \includegraphics[width=\linewidth]{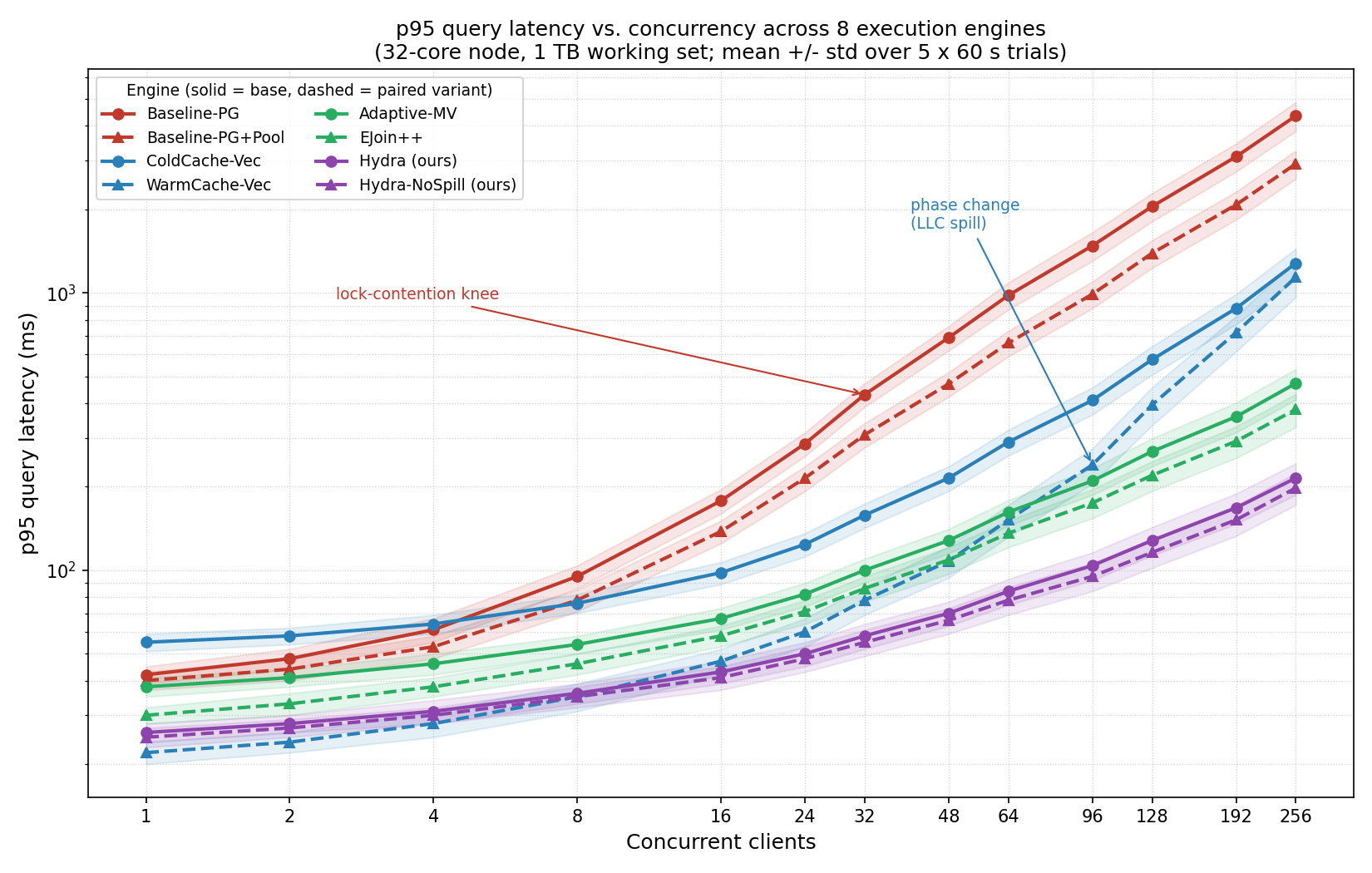}
  \end{minipage}\hfill
  \begin{minipage}{0.495\textwidth}\centering
    \includegraphics[width=\linewidth]{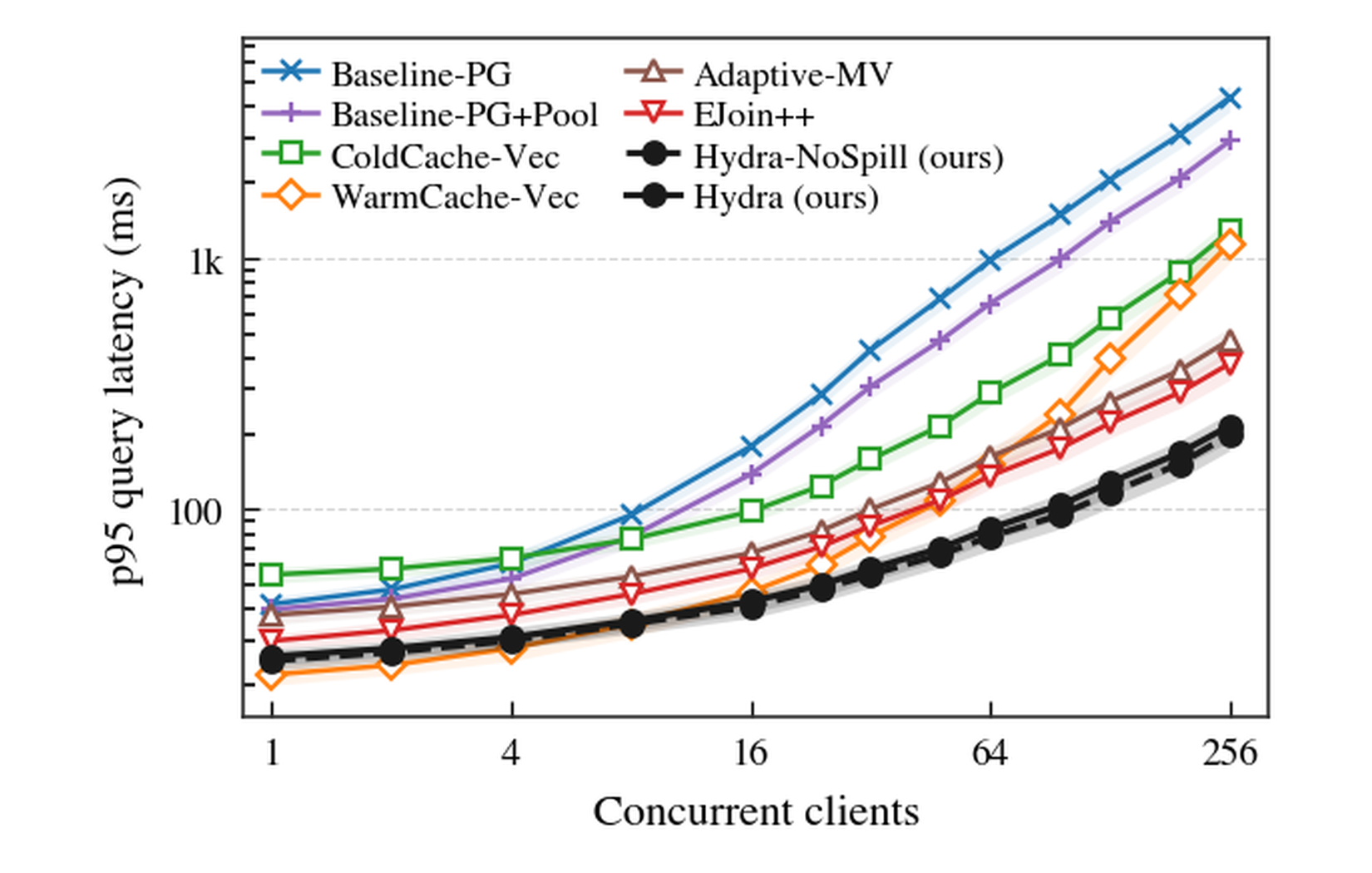}
  \end{minipage}
  \\[4pt]
  \begin{minipage}{0.495\textwidth}\centering
    \includegraphics[width=\linewidth]{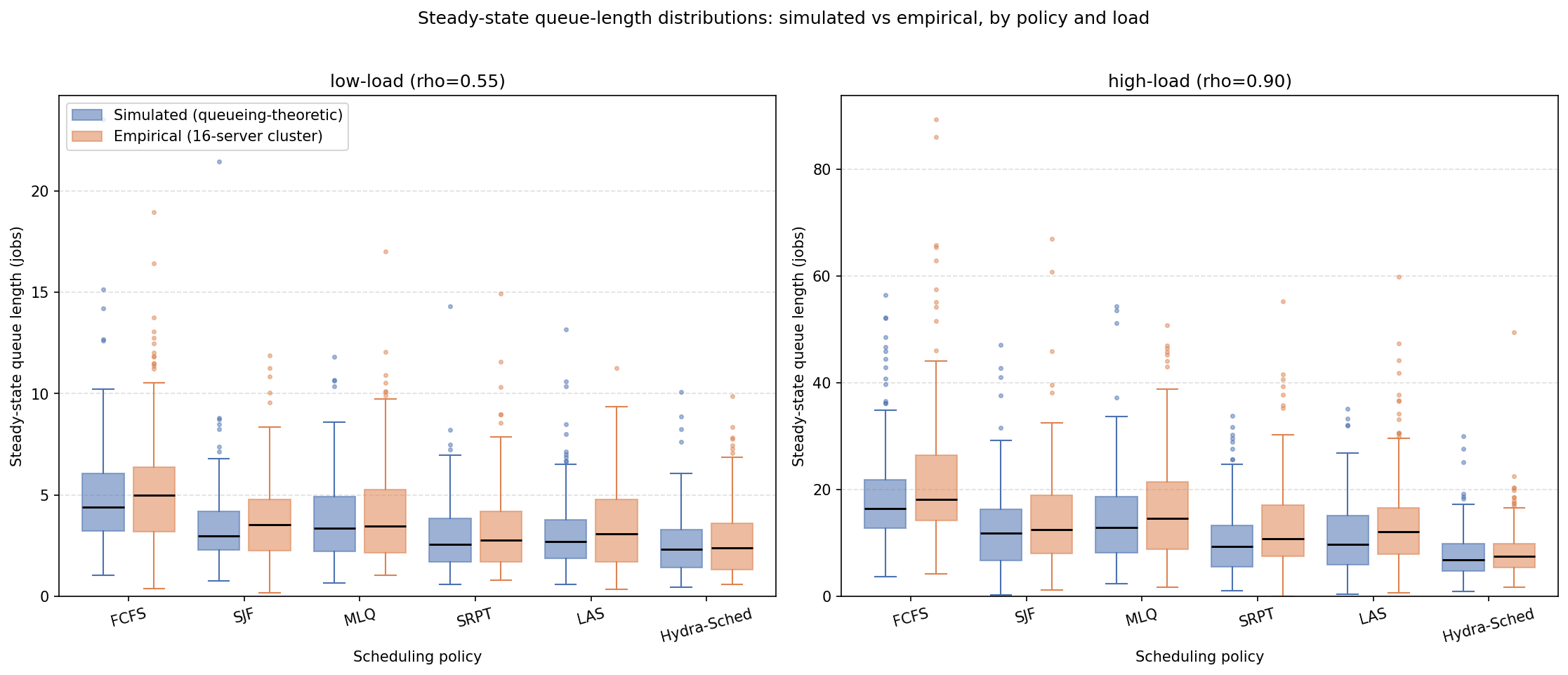}
  \end{minipage}\hfill
  \begin{minipage}{0.495\textwidth}\centering
    \includegraphics[width=\linewidth]{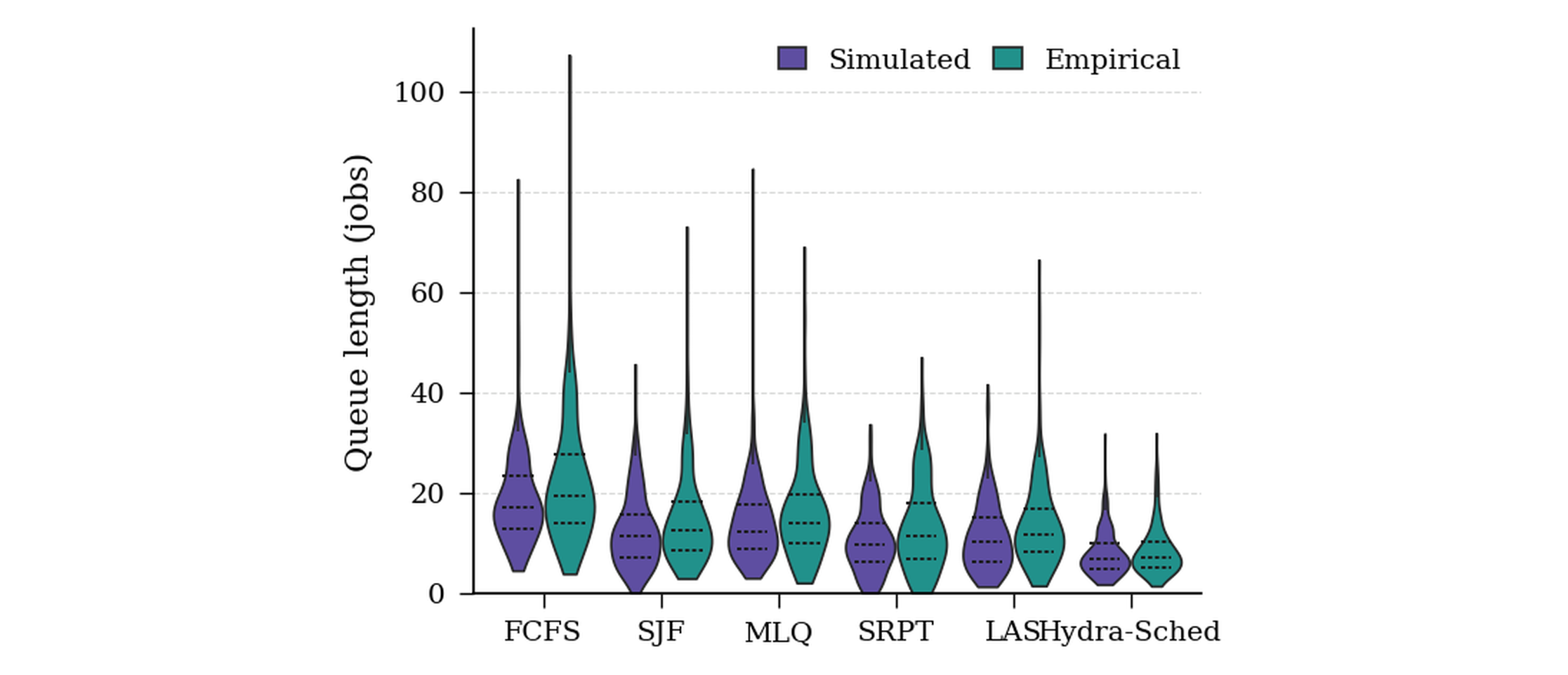}
  \end{minipage}
  \caption{Three style cases (rows, top to bottom: Case~1
  grouped-bar ablation; Case~2 scaling line chart; Case~3
  paired-condition distribution comparison). \textbf{Left
  column:} the chart a baseline coding LLM (the default harness,
  no style skill loaded) produces from the same intent and data.
  \textbf{Right column:} the chart \system{} produces with the
  distilled VLDB style skill loaded. The baseline outputs show
  recurring publication-quality issues: off-venue palettes, missing
  hatching or marker discipline, tick labels and legends that
  become unreadable once the chart is rendered at column width,
  and inconsistent error-bar styling. The right-column variants
  inherit the venue's visual conventions and remain legible at
  the same column width.}
  \label{fig:style_cases}
\end{figure*}

A new venue or a lab-specific palette is captured by a new skill
file rather than by retraining.
Figure~\ref{fig:style_cases} visualises the effect across three
chart types: a grouped-bar ablation, a paired-marker scaling line
chart, and a paired-condition distribution comparison. In each
row we compare the chart Claude writes from the same intent and
data with (right) and without (left) the distilled VLDB style
skill loaded into its context.

\subsection{Deployment-aware render loop}
\label{sec:method:deploy}

\noindent\textbf{Key insight.} Charts are produced in isolation
but consumed inside a paper column. We close this gap by
compiling the chart \emph{inside} the target \LaTeX{} template,
which serves two purposes. First, the system observes
layout-level errors (column overflow, illegible fonts,
aspect-ratio drift) that no standalone \texttt{matplotlib} render
reveals, and feeds them back into a code-revision loop. Second,
the chart now lives in the same rendering context in which the
author will eventually edit it
(\S\ref{sec:method:edit}); a single in-context adjustment
produces the camera-ready result, instead of an iteration cycle
between an isolated preview and the target paper.

\begin{figure}[h]
  \centering
  \includegraphics[width=\linewidth]{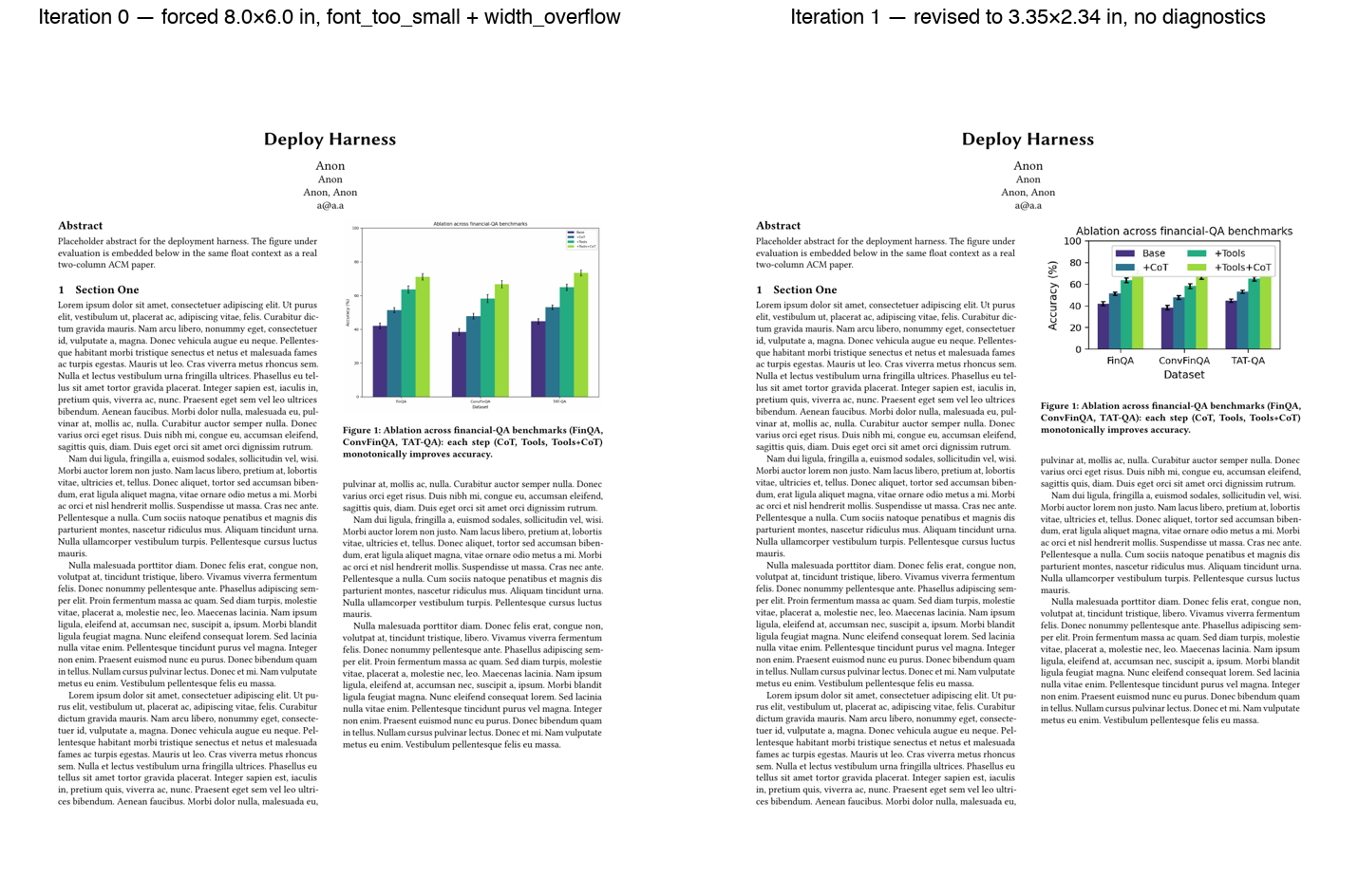}
  \caption{Deploy-loop trace, iteration~0 (left) and iteration~1
  (right). The forced 8$\times$6\,in chart trips a width violation
  (8.0\,in vs.\ 3.35\,in column) and a font-size violation
  (effective $\approx 4.2$\,pt tick labels). One revision
  regenerates the figure at 3.35$\times$2.34\,in; all checks pass.}
  \label{fig:deploy}
\end{figure}

\noindent\textbf{Worked example.} An oversized 8$\times$6\,inch chart
compiled inside a 3.35-inch single column trips a width violation
and a font-size violation. One revision regenerates the figure at
$3.35\times2.34$\,in and the loop converges
(Figure~\ref{fig:deploy}).

\noindent\textbf{The loop.} The harness compiles the generated
chart inside a template that mirrors the target paper's column
width, font budget, and float style. After the compile, it
measures the rendered figure's bounding box and produces structured
diagnostics for any layout violation: compile failures, width or
font issues, aspect-ratio drift, and caption overflow. Each
diagnostic is converted into a concrete code-edit instruction (for
instance, ``increase tick-label font size to at least 6\,pt'') and
fed back into the next round of code generation. The loop
terminates on convergence, iteration cap, compile error, or model
failure.

\subsection{Multimodal edit layer}
\label{sec:method:edit}

\noindent\textbf{Key insight.} Re-prompting an LLM with desired
changes has two limits. First, free-text instructions like
``move the legend a bit'' rarely converge on what the author
meant, because the preference is geometric, not semantic. Second,
the LLM never sees the rendered chart; it sees only the code, so
each edit is a hypothesis about the visual outcome, and the
divergence compounds across
iterations~\cite{matplotagent2024,visshepherd2025}. We provide a
multimodal input path instead: the author manipulates chart
elements directly inside the same \LaTeX{} environment used for
deployment, so what they see during editing is what the
camera-ready paper will show.

\begin{figure}[t]
  \centering
  \includegraphics[width=\linewidth]{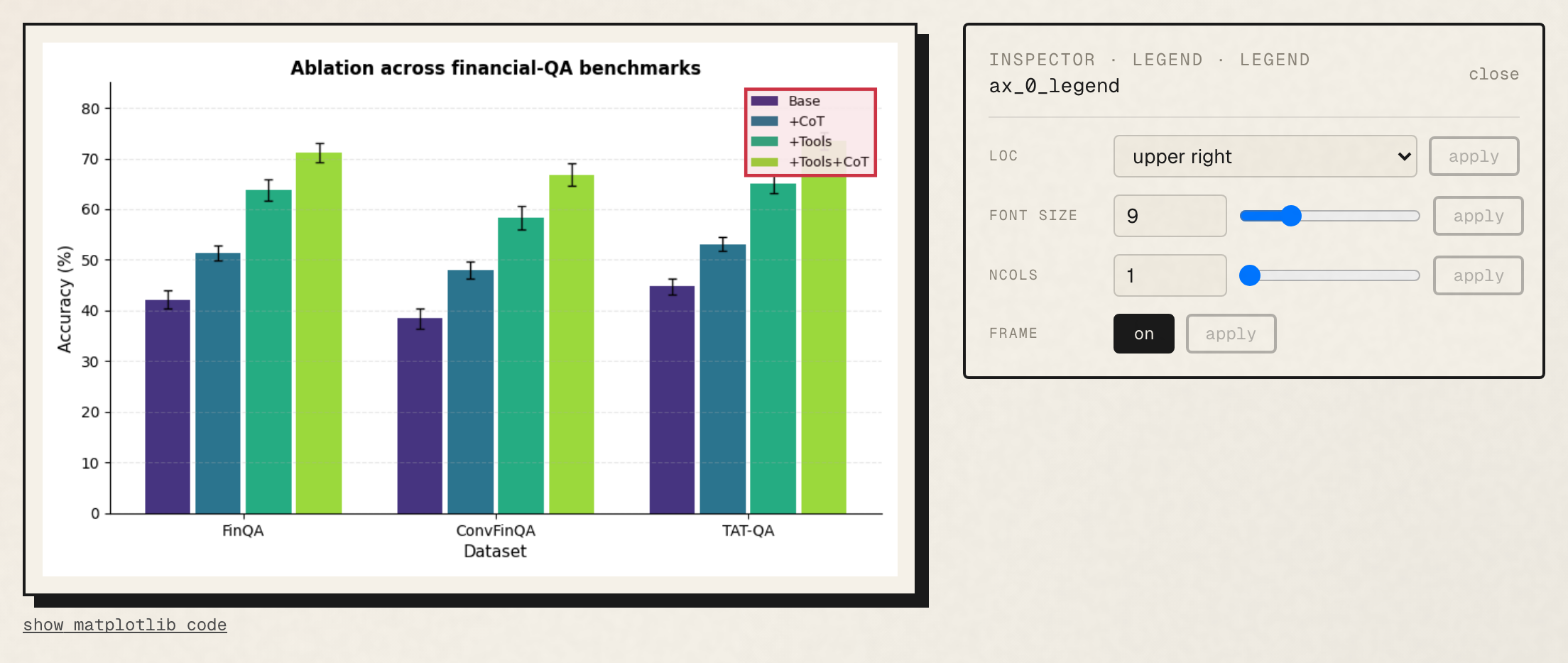}
  \caption{The edit layer, captured from the live web interface.
  (a) Element targets are overlaid on the rendered chart; the
  legend is currently selected (red highlight). (b) The property
  inspector for the selected legend exposes its addressable
  controls (location, font size, column count, frame); each
  control maps to a concrete \texttt{matplotlib} mutation that is
  re-applied through the deployment harness.}
  \label{fig:edit}
\end{figure}

\noindent\textbf{The interface.} The system executes the generated
code in a sandbox, walks the \texttt{matplotlib} \texttt{Figure}
object tree, and assigns every element (axis label, legend, line,
patch) a stable identifier and bounding box. A web interface
overlays click-targets on the rendered figure; selecting an
element opens a contextual inspector whose controls depend on the
element kind: a legend exposes location, column count, font size,
and frame toggle; a line exposes colour, width, style, and marker;
a text element exposes content, size, and colour. Each control
maps to a concrete code mutation (e.g.\
\texttt{ax.legend(loc='upper left', ncol=2)}); because edits
target the source code rather than a rendered bitmap, every
change is reproducible and composes with automated deployment
fixes (\autoref{fig:edit}).

\section{Case Study and Early Results}
\label{sec:case}

We walk three chart cases through the \system{} pipeline. Each
case starts with an intent and data, runs through skill-prefixed
codegen, and lands in a target VLDB \texttt{acmart} column. The
three cases sweep chart types common to systems papers: a
grouped-bar ablation (Case~1), a paired-marker scaling line chart
(Case~2), and a paired-condition distribution comparison
(Case~3). All three are shown side by side as the rows of
Figure~\ref{fig:style_cases}, baseline on the left and \system{}
output on the right.

\subsection{Three cases}

\paragraph{Case~1: grouped-bar ablation.} The input compares
MRR with vs.\ without fine-tuning across five datasets
(ICEWS~14, ICEWS~05-15, YAGO~11k, Wikidata~12k, GDELT). The
default Claude Code output (Figure~\ref{fig:style_cases}, row~1
left) reads more like an AI-paper rendering than a systems-paper
chart: a saturated palette, sans-serif labels, no hatching, and
tick fonts that shrink below legibility once the chart is dropped
into a single column. The \system{} variant (row~1 right)
inherits the venue's serif typography, muted Family-A palette,
hatched bars, and a frameless compact legend.

\paragraph{Case~2: paired-marker scaling line chart.} The intent
asks how p95 query latency for eight execution engines changes as
concurrent clients scale from 1 to 256. The baseline (row~2 left)
draws eight equally weighted lines that compete for attention and
overflow tick labels at column width. The \system{} variant
(row~2 right) pairs the two proposed variants (filled black)
against six baselines (open coloured markers), uses log-log axes
with mean-$\pm$-std error bands, and selects a column-aware
\texttt{figsize}.

\paragraph{Case~3: paired-condition distribution comparison.} The
intent compares queue-length distributions from a simulator
against empirical traces under high load for six scheduling
policies. The baseline (row~3 left) emits a default boxplot per
policy that obscures distribution shape and tail behaviour. The
\system{} variant (row~3 right) renders paired violins per policy
(sim purple, emp teal) with embedded quartile ticks, surfacing
the sim-vs-emp gap directly.

Together the three cases share a single observation: a default
coding-LLM harness produces plausible \texttt{matplotlib} code,
yet the resulting figure misses both the venue's visual
conventions and the column-width legibility budget. The
\system{} harness fixes both: the style skill encodes the right
defaults and the deployment loop keeps the result legible.

\subsection{Observations}

\begin{itemize}[leftmargin=*, itemsep=2pt]
  \item \textbf{Default code-LLM output is plausible but not
  publication-ready.} Across all three cases the baseline
  chart compiles cleanly yet fails both the venue conventions
  and the column-width legibility test
  (Figure~\ref{fig:style_cases}, left column). The \system{}
  variant addresses both in a single pass.
  \item \textbf{Constraint violations are diagnosable.} When
  upstream code does fail a layout check, the deployment loop
  surfaces a named diagnostic with measured offending values
  (font size, width, aspect ratio) rather than a free-text
  vision-model critique, so the next codegen iteration knows
  what to fix.
  \item \textbf{Geometric edits cost zero LLM tokens.} Author
  edits in the inspector go through AST mutation; the
  alternative would be a full re-prompt of $\sim$3000 tokens to
  describe the chart and the desired change.
  \item \textbf{Skill and deployment are a co-design.} When the
  skill already encodes a column-aware figure budget, the
  deployment loop converges immediately; it fires only when
  upstream code violates a constraint.
\end{itemize}

\subsection{User study}
\label{sec:case:user_study}

To complement the case studies with feedback from real authors,
we ran a small user study with four computer science researchers
familiar with academic publishing. Participants rated \system{}
against a baseline coding LLM on a 5-point Likert scale across
five questions, organised as one overall question and three
component-specific groups:

\begin{itemize}[leftmargin=*, itemsep=2pt]
  \item \textbf{Overall.} Q1: would you continue using the system
  for future submissions?
  \item \textbf{Style component.} Q2: does the figure match
  top-venue VLDB visual conventions? Q3: how aesthetically
  appealing is the figure overall?
  \item \textbf{Deploy component.} Q4: are tick labels, legends,
  and titles readable at column width?
  \item \textbf{Edit component.} Q5: how easy and precise is it to
  apply small geometric edits to the figure?
\end{itemize}

Q1--Q4 were rated by direct inspection of the three cases in
\S\ref{sec:case} (the same figures the reader sees here),
comparing each \system{} output against a no-skill baseline
rendering. Q5 was rated after a short live demo of the edit
interface in which each participant performed a few precise edits
on a \system{} output in the browser.

\begin{figure}[t]
  \centering
  \includegraphics[width=\linewidth]{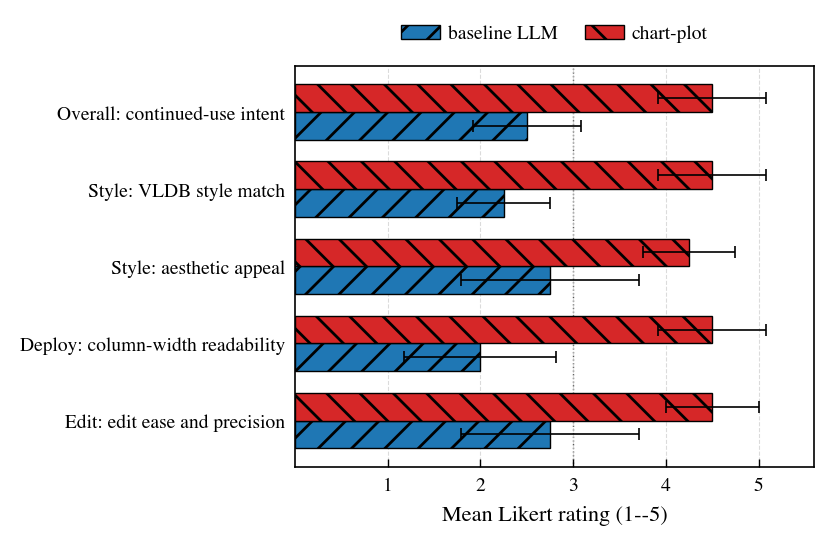}
  \caption{User study results (N=4 computer science researchers).
  Mean Likert rating per question with one-standard-deviation
  error bars, comparing a baseline coding LLM against \system{}
  across five questions in four groups (Overall, Style, Deploy,
  Edit). This figure was itself generated end-to-end by the
  \system{} skill under the same single-column \texttt{acmart}
  target as the rest of the paper.}
  \label{fig:user_study}
\end{figure}

\autoref{fig:user_study} summarises the results. \system{} lifts
mean ratings by 1.50--2.50 Likert points across all five
questions; the largest gaps fall on column-width readability
(Q4, 4.5 vs.\ 2.00) and continued-use intent (Q1, 4.5 vs.\ 2.50),
and the smallest on aesthetic appeal (Q3, 4.25 vs.\ 2.75). The
edit dimension (Q5) lands at 4.5 vs.\ 2.75 once participants have
spent a few minutes inside the click-target inspector.

\paragraph{Interview insights.} A short post-task interview with
each participant surfaced two recurring threads worth
documenting. First, participants converged on a hybrid editing
pattern: chat-style natural-language instructions worked best
for \emph{large} changes such as rewriting an annotation or
restructuring a multi-axis layout, while the click-target
inspector worked best for \emph{small, precise} geometric edits.
We accordingly recommend a hybrid multimodal editing workflow
that combines both. Second, participants observed that \system{}
ships two harness environments (the standalone implementation
used for the demo and a Claude Code skill packaging) and asked
for parity with other coding agents such as OpenAI Codex;
broadening the supported set of agent harnesses is the most
direct lever for community adoption.

\section{Discussion and Conclusion}
\label{sec:discussion}

We presented \system{}, an agentic harness that closes the last
mile of academic chart generation through three layers: a
style-aware code generator, a deployment-aware render loop, and a
multimodal edit layer.

What \system{} distils today is the visual idiom of one broad
community (computer-science papers). Every academic field
carries its own conventions that a general-purpose LLM does not
reproduce reliably from training data alone; financial charts,
for example, annotate inflection points, regime changes, and
event windows with ad-hoc text labels and shaded
spans~\cite{luo2025finsignal,jiang2025autoMA}. Our next step is
an open-source library of expert-distilled style skills across
additional domains, packaged in the same skill format introduced
in \S\ref{sec:method:style}, so that academic chart generation
does not stop at one community's gold standard.

\bibliographystyle{ACM-Reference-Format}
\bibliography{references}

\end{document}